\documentclass[a4paper]{report}
\usepackage[utf8]{inputenc}
\usepackage[T1]{fontenc}
\usepackage{RJournal}
\usepackage{amsmath,amssymb,array}
\usepackage{booktabs}

\begin{document}

\sectionhead{Contributed research article}
\volume{XX}
\volnumber{YY}
\year{20ZZ}
\month{AAAA}

\begin{article}
\title{Advanced Bayesian Multilevel Modeling with the R Package brms}
\author{Paul-Christian B\"urkner}

\maketitle

\abstract{
The \pkg{brms} package allows R users to easily specify a wide range of Bayesian single-level and multilevel models, which are fitted with the probabilistic programming language Stan behind the scenes. Several response distributions are supported, of which all parameters (e.g., location, scale, and shape) can be predicted at the same time thus allowing for distributional regression. Non-linear relationships may be specified using non-linear predictor terms or semi-parametric approaches such as splines or Gaussian processes. To make all of these modeling options possible in a multilevel framework, \pkg{brms} provides an intuitive and powerful formula syntax, which extends the well known formula syntax of \pkg{lme4}. The purpose of the present paper is to introduce this syntax in detail and to demonstrate its usefulness with four examples, each showing other relevant aspects of the syntax.
}

\section{Introduction}

Multilevel models (MLMs) offer great flexibility for researchers across sciences \citep{brown2015, demidenko2013, gelmanMLM2006, pinheiro2006}. They allow modeling of data measured on different levels at the same time -- for instance data of students nested within classes and schools -- thus taking complex dependency structures into account. It is not surprising that many packages for R have been developed to fit MLMs. Usually, however, the functionality of these implementations is limited insofar as it is only possible to predict the mean of the response distribution. Other parameters of the response distribution, such as the residual standard deviation in linear models, are assumed constant across observations, which may be violated in many applications. Accordingly, it is desirable to allow for prediction of \emph{all} response parameters at the same time. Models doing exactly that are often referred to as \emph{distributional models} or more verbosely \emph{models for location, scale and shape} \citep{rigby2005}. Another limitation of basic MLMs is that they only allow for linear predictor terms. While linear predictor terms already offer good flexibility, they are of limited use when relationships are inherently non-linear. Such non-linearity can be handled in at least two ways: (1) by fully specifying a non-linear predictor term with corresponding parameters each of which can be predicted using MLMs \citep{lindstrom1990}, or (2) estimating the form of the non-linear relationship on the fly using splines \citep{wood2004} or Gaussian processes \citep{rasmussen2006}. The former are often simply called \emph{non-linear models}, while models applying splines are referred to as \emph{generalized additive models} (GAMs; \citeauthor{hastie1990}, \citeyear{hastie1990}). 

Combining all of these modeling options into one framework is a complex task, both conceptually and with regard to model fitting. Maximum likelihood methods, which are typically applied in classical 'frequentist' statistics, can reach their limits at some point and fully Bayesian methods become the go-to solutions to fit such complex models \citep{gelman2014}. In addition to being more flexible, the Bayesian framework comes with other advantages, for instance, the ability to derive probability statements for every quantity of interest or explicitly incorporating prior knowledge about parameters into the model. The former is particularly relevant in non-linear models, for which classical approaches struggle more often than not in propagating all the uncertainty in the parameter estimates to non-linear functions such as out-of-sample predictions.

Possibly the most powerful program for performing full Bayesian inference available to date is Stan \citep{stanM2017, carpenter2017}. It implements Hamiltonian Monte Carlo \citep{duane1987, neal2011, betancourt2014} and its extension, the No-U-Turn (NUTS) Sampler  \citep{hoffman2014}. These algorithms converge much more quickly than other Markov-Chain Monte-Carlo (MCMC) algorithms especially for high-dimensional models \citep{hoffman2014, betancourt2014, betancourt2017}. An excellent non-mathematical introduction to Hamiltonian Monte Carlo can be found in \citet{betancourt2017}.

Stan comes with its own programming language, allowing for great modeling flexibility \cite{stanM2017, carpenter2017}). Many researchers may still be hesitent to use Stan directly, as every model has to be written, debugged and possibly also optimized. This may be a time-consuming and error-prone process even for researchers familiar with Bayesian inference. The \CRANpkg{brms} package \citet{buerkner2017}, presented in this paper, aims to remove these hurdles for a wide range of regression models by allowing the user to benefit from the merits of Stan by using extended \CRANpkg{lme4}-like \citep{bates2015} formula syntax, with which many R users are familiar with. It offers much more than writing efficient and human-readable Stan code: \pkg{brms} comes with many post-processing and visualization functions, for instance to perform posterior predictive checks, leave-one-out cross-validation, visualization of estimated effects, and prediction of new data. The overarching aim is to have one general framework for regression modeling, which offers everything required to successfully apply regression models to complex data. To date, it already replaces and extends the functionality of dozens of other R packages, each of which is restricted to specific regression models\footnote{Unfortunately, due to the implementation via Stan, it is not easily possible for users to define their own response distributions and run them via \pkg{brms}. If you feel that a response distribution is missing in \pkg{brms}, please open an issue on GitHub (\url{https://github.com/paul-buerkner/brms}).}.

The purpose of the present article is to provide an introduction of the advanced multilevel formula syntax implemented in \pkg{brms}, which allows to fit a wide and growing range of  non-linear distributional multilevel models. A general overview of the package is already given in \citet{buerkner2017}. Accordingly, the present article focuses on more recent developments. We begin by explaining the underlying structure of distributional models. Next, the formula syntax of \pkg{lme4} and its extensions implemented in \pkg{brms} are explained. Four examples that demonstrate the use of the new syntax are discussed in detail. We end by describing future plans for extending the package. 

\section{Model description}
\label{model}

The core of models implemented in \pkg{brms} is the prediction of the response $y$ through predicting all parameters $\theta_p$ of the response distribution $D$, which is also called the model \code{family} in many R packages. We write 
$$y_i \sim D(\theta_{1i}, \theta_{2i}, ...)$$
to stress the dependency on the $i\textsuperscript{th}$ observation. Every parameter $\theta_p$ may be regressed on its own predictor term $\eta_p$ transformed by the inverse link function $f_p$ that is $\theta_{pi} = f_p(\eta_{pi})$\footnote{A parameter can also be assumed constant across observations so that a linear predictor is not required.}. Such models are typically refered to as \emph{distributional models}\footnote{The models described in \citet{buerkner2017} are a sub-class of the here described models.}. Details about the parameterization of each \code{family} are given in \code{vignette("brms\_families")}.

Suppressing the index $p$ for simplicity, a predictor term $\eta$ can generally be written as
$$
\eta = \mathbf{X} \beta + \mathbf{Z} u + \sum_{k = 1}^K s_k(x_k)
$$
In this equation, $\beta$ and $u$ are the coefficients at population-level and group-level respectively and $\mathbf{X}, \mathbf{Z}$ are the corresponding design matrices. The terms $s_k(x_k)$ symbolize optional smooth functions of unspecified form based on covariates $x_k$ fitted via splines (see \citet{wood2011} for the underlying implementation in the \CRANpkg{mgcv} package) or Gaussian processes \citep{williams1996}. The response $y$ as well as $\mathbf{X}$, $\mathbf{Z}$, and $x_k$ make up the data, whereas $\beta$, $u$, and the smooth functions $s_k$ are the model parameters being estimated. The coefficients $\beta$ and $u$ may be more commonly known as fixed and random effects, but I avoid theses terms following the recommendations of \citet{gelmanMLM2006}. Details about prior distributions of $\beta$ and $u$ can be found in \citet{buerkner2017} and under \code{help("set\_prior")}.
 
As an alternative to the strictly additive formulation described above, predictor terms may also have any form specifiable in Stan. We call it a \emph{non-linear} predictor and write 
$$\eta = f(c_1, c_2, ..., \phi_1, \phi_2, ...)$$
The structure of the function $f$ is given by the user, $c_r$ are known or observed covariates, and $\phi_s$ are non-linear parameters each having its own linear predictor term $\eta_{\phi_s}$ of the form specified above. In fact, we should think of non-linear parameters as placeholders for linear predictor terms rather than as parameters themselves. A frequentist implementation of such models, which inspired the non-linear syntax in \pkg{brms}, can be found in the \CRANpkg{nlme} package \citep{nlme2016}. 

\section{Extended multilevel formula syntax}
\label{formula_syntax}

The formula syntax applied in \pkg{brms} builds upon the syntax of the R package \pkg{lme4} \citep{bates2015}. First, we will briefly explain the \pkg{lme4} syntax used to specify multilevel models and then introduce certain extensions that allow to specify much more complicated models in \pkg{brms}. An \pkg{lme4} formula has the general form

\begin{example}
response ~ pterms + (gterms | group)
\end{example}
The \code{pterms} part contains the population-level effects that are assumed to be the same across obervations. The \code{gterms} part contains so called group-level effects that are assumed to vary accross grouping variables specified in \code{group}. Multiple grouping factors each with multiple group-level effects are possible.  Usually, \code{group} contains only a single variable name pointing to a factor, but you may also use \code{g1:g2} or \code{g1/g2}, if both \code{g1} and \code{g2} are suitable grouping factors. The \code{:} operator creates a new grouping factor that consists of the combined levels of \code{g1} and \code{g2} (you could think of this as pasting the levels of both factors together). The \code{/} operator indicates nested grouping structures and expands one grouping factor into two or more when using multiple \code{/} within one term. If, for instance, you write \code{(1 | g1/g2)}, it will be expanded to \code{(1 | g1) + (1 | g1:g2)}. Instead of \code{|} you may use \code{||} in grouping terms to prevent group-level correlations from being modeled. This may be useful in particular when modeling so many group-level effects that convergence of the fitting algorithms becomes an issue due to model complexity. One limitation of the \code{||} operator in \pkg{lme4} is that it only splits up terms so that columns of the design matrix originating from the same term are still modeled as correlated (e.g., when coding a categorical predictor; see the \code{mixed} function of the \CRANpkg{afex} package by \citet{afex2015} for a way to avoid this behavior). 

While intuitive and visually appealing, the classic \pkg{lme4} syntax is not flexible enough to allow for specifying the more complex models supported by \pkg{brms}. In non-linear or distributional models, for instance, multiple parameters are predicted, each having their own population and group-level effects. Hence, multiple formulas are necessary to specify such models\footnote{Actually, it is possible to specify multiple model parts within one formula using interactions terms for instance as implemented in \CRANpkg{MCMCglmm} \citep{hadfield2010}. However, this syntax is limited in flexibility and requires a rather deep understanding of the way R parses formulas, thus often being confusing to users.}. Then, however, specifying group-level effects of the same grouping factor to be correlated \emph{across} formulas becomes complicated. The solution implemented in \pkg{brms} (and currently unique to it) is to expand the \code{|} operator into \code{|<ID>|}, where \code{<ID>} can be any value. Group-level terms with the same \code{ID} will then be modeled as correlated if they share same grouping factor(s)\footnote{It might even be further extended to \code{|fun(<ID>)|}, where \code{fun} defines the type of correlation structure, defaulting to unstructured that is estimating the full correlation matrix. The \code{fun} argument is not yet supported by \pkg{brms} but could be supported in the future if other correlation structures, such as compound symmetry or Toeplitz, turn out to have reasonable practical applications and effective implementations in Stan.}.  For instance, if the terms \code{(x1|ID|g1)} and \code{(x2|ID|g1)} appear somewhere in the same or different formulas passed to \pkg{brms}, they will be modeled as correlated.

Further extensions of the classical \pkg{lme4} syntax refer to the \code{group} part. It is rather limited in its flexibility since only variable names combined by \code{:} or \code{/} are supported. We propose two extensions of this syntax: Firstly, \code{group} can generally be split up in its terms so that, say, \code{(1 | g1 + g2)} is expanded to \code{(1 | g1) + (1 | g2)}. This is fully consistent with the way \code{/} is handled so it provides a natural generalization to the existing syntax. Secondly, there are some special grouping structures that cannot be expressed by simply combining grouping variables. For instance, multi-membership models cannot be expressed this way. To overcome this limitation, we propose wrapping terms in \code{group} within special functions that allow specifying alternative grouping structures: \code{(gterms | fun(group))}. In \pkg{brms}, there are currently two such functions implemented, namely \code{gr} for the default behavior and \code{mm} for multi-membership terms. To be compatible with the original syntax and to keep formulas short, \code{gr} is automatically added internally if none of these functions is specified.

While some non-linear relationships, such as quadratic relationships, can be expressed within the basic R formula syntax, other more complicated ones cannot. For this reason, it is possible in \pkg{brms} to fully specify non-linear predictor terms similar to how it is done in \pkg{nlme}, but fully compatible with the extended multilevel syntax described above. Suppose, for instance, we want to model the non-linear growth curve 
$$
y = b_1 (1 - \exp(-(x / b_2)^{b_3})
$$
between $y$ and $x$ with parameters $b_1$, $b_2$, and $b_3$ (see Example 3 in this paper for an implementation of this model with real data). Furthermore, we want all three parameters to vary by a grouping variable $g$ and model those group-level effects as correlated. Additionally $b_1$ should be predicted by a covariate $z$. We can express this in \pkg{brms} using multiple formulas, one for the non-linear model itself and one per non-linear parameter:
\begin{example}
y ~ b1 * (1 - exp(-(x / b2) ^ b3)
b1 ~ z + (1|ID|g)
b2 ~ (1|ID|g)
b3 ~ (1|ID|g)
\end{example}
The first formula will not be evaluated using standard R formula parsing, but instead taken literally. In contrast, the formulas for the non-linear parameters will be evaluated in the usual way and are compatible with all terms supported by \pkg{brms}. Note that we have used the above described ID-syntax to model group-level effects as correlated across formulas.

There are other syntax extensions implemented in \pkg{brms} that do not directly target grouping terms. Firstly, there are terms formally included in the \code{pterms} part that are handled separately. The most prominent examples are smooth terms specified through the \code{s} and \code{t2} functions of the \pkg{mgcv} package \citep{wood2011}. Other examples are category specific effects \code{cs}, monotonic effects \code{mo}, noise-free effects \code{me}, or Gaussian process terms \code{gp}. The former is explained in \citet{buerkner2017}, while the latter three are documented in \code{help(brmsformula)}. Internally, these terms are extracted from \code{pterms} and not included in the construction of the population-level design matrix. Secondly, making use of the fact that \code{|} is unused on the left-hand side of $\sim$ in formula, additional information on the response variable may be specified via 
\begin{example}
response | aterms ~ <predictor terms>
\end{example}
The \code{aterms} part may contain multiple terms of the form \code{fun(<variable>)} separated by \code{+} each providing special information on the response variable. This allows among others to weight observations, provide known standard errors for meta-analysis, or model censored or truncated data. 
As it is not the main topic of the present paper, we refer to \code{help("brmsformula")} and \code{help("addition-terms")} for more details.

To set up the model formulas and combine them into one object, \pkg{brms} defines the \code{brmsformula} (or short \code{bf}) function. Its output can then be passed to the \code{parse\_bf} function, which splits up the formulas in separate parts and prepares them for the generation of design matrices and related data. Other packages may re-use these functions in their own routines making it easier to offer support for the above described multilevel syntax.

\section{Examples}

The idea of \pkg{brms} is to provide one unified framework for  multilevel regression models in R. As such, the above described formula syntax in all of its variations can be applied in combination with all response distributions supported by \pkg{brms} (currently about 35 response distributions are supported; see \code{help("brmsfamily")} and \code{vignette("brms\_families")} for an overview).

In this section, we will discuss four examples in detail, each focusing on certain aspects of the syntax. They are chosen to provide a broad overview of the modeling options. The first is about the number of fish caught be different groups of people. It does not actually contain any multilevel structure, but helps in understanding how to set up formulas for different model parts. The second example is about housing rents in Munich. We model the data using splines and a distributional regression approach. The third example is about cumulative insurance loss payments across several years, which is fitted using a rather complex non-linear multilevel model. Finally, the fourth example is about the performance of school children, who change school during the year, thus requiring a multi-membership model.

Despite not being covered in the four examples, there are a few more modeling options that we want to briefly describe. First, \pkg{brms} allows fitting so called phylogenetic models. These models are relevant in evolutionary biology when data of many species are analyzed at the same time. Species are not independent as they come from the same phylogenetic tree, implying that different levels of the same grouping-factor (i.e., species) are likely correlated. There is a whole vignette dedicated to this topic, which can be found via \code{vignette("brms\_phylogenetics")}. Second, there is a canonical way to handle ordinal predictors, without falsely assuming they are either categorical or continuous. We call them monotonic effects and discuss them in \code{vignette("brms\_monotonic")}. Last but not least, it is possible to account for measurement error in both response and predictor variables. This is often ignored in applied regression modeling \citep{westfall2016}, although measurement error is very common in all scientific fields making use of observational data. There is no vignette yet covering this topic, but one will be added in the future. In the meantime, \code{help("brmsformula")} is the best place to start learning about such models as well as about other details of the \pkg{brms} formula syntax. 

\subsection{Example 1: Catching fish}

An important application of the distributional regression framework of \pkg{brms} are so called zero-inflated and hurdle models. These models are helpful whenever there are more zeros in the response variable than one would naturally expect. Here, we consider an example dealing with the number of fish caught by various groups of people. On the UCLA website (\url{https://stats.idre.ucla.edu/stata/dae/zero-inflated-poisson-regression}), the data are described as follows: ``The state wildlife biologists want to model how many fish are being caught by fishermen at a state park. Visitors are asked how long they stayed, how many people were in the group, were there children in the group and how many fish were caught. Some visitors do not fish, but there is no data on whether a person fished or not. Some visitors who did fish did not catch any fish so there are excess zeros in the data because of the people that did not fish.''

\begin{example}
zinb <- read.csv("http://stats.idre.ucla.edu/stat/data/fish.csv")
zinb$camper <- factor(zinb$camper, labels = c("no", "yes"))
head(zinb)
\end{example}

\begin{example}
  nofish livebait camper persons child         xb         zg count
1      1        0     no       1     0 -0.8963146  3.0504048     0
2      0        1    yes       1     0 -0.5583450  1.7461489     0
3      0        1     no       1     0 -0.4017310  0.2799389     0
4      0        1    yes       2     1 -0.9562981 -0.6015257     0
5      0        1     no       1     0  0.4368910  0.5277091     1
6      0        1    yes       4     2  1.3944855 -0.7075348     0
\end{example}
As predictors we choose the number of people per group, the number of children, as well as whether or not the group consists of campers. Many groups may not catch any fish just because they do not try and so we fit a zero-inflated Poisson model. For now, we assume a constant zero-inflation probability across observations.

\begin{example}
fit_zinb1 <- brm(count ~ persons + child + camper, data = zinb, 
                 family = zero_inflated_poisson("log"))
\end{example}
The model is readily summarized via

\begin{example}
summary(fit_zinb1)
\end{example}

\begin{example}
 Family: zero_inflated_poisson (log) 
Formula: count ~ persons + child + camper 
   Data: zinb (Number of observations: 250) 
Samples: 4 chains, each with iter = 2000; warmup = 1000; thin = 1; 
         total post-warmup samples = 4000
   WAIC: Not computed
 
Population-Level Effects: 
          Estimate Est.Error l-95
Intercept    -1.01      0.17    -1.34    -0.67       2171    1
persons       0.87      0.04     0.79     0.96       2188    1
child        -1.36      0.09    -1.55    -1.18       1790    1
camper        0.80      0.09     0.62     0.98       2950    1

Family Specific Parameters: 
   Estimate Est.Error l-95
zi     0.41      0.04     0.32     0.49       2409    1

Samples were drawn using sampling(NUTS). For each parameter, Eff.Sample 
is a crude measure of effective sample size, and Rhat is the potential 
scale reduction factor on split chains (at convergence, Rhat = 1).
\end{example}
A graphical summary is available through

\begin{example}
marginal_effects(fit_zinb1)
\end{example}
\begin{figure}[ht]
  \centering
  \includegraphics[width=0.99\textwidth,keepaspectratio]{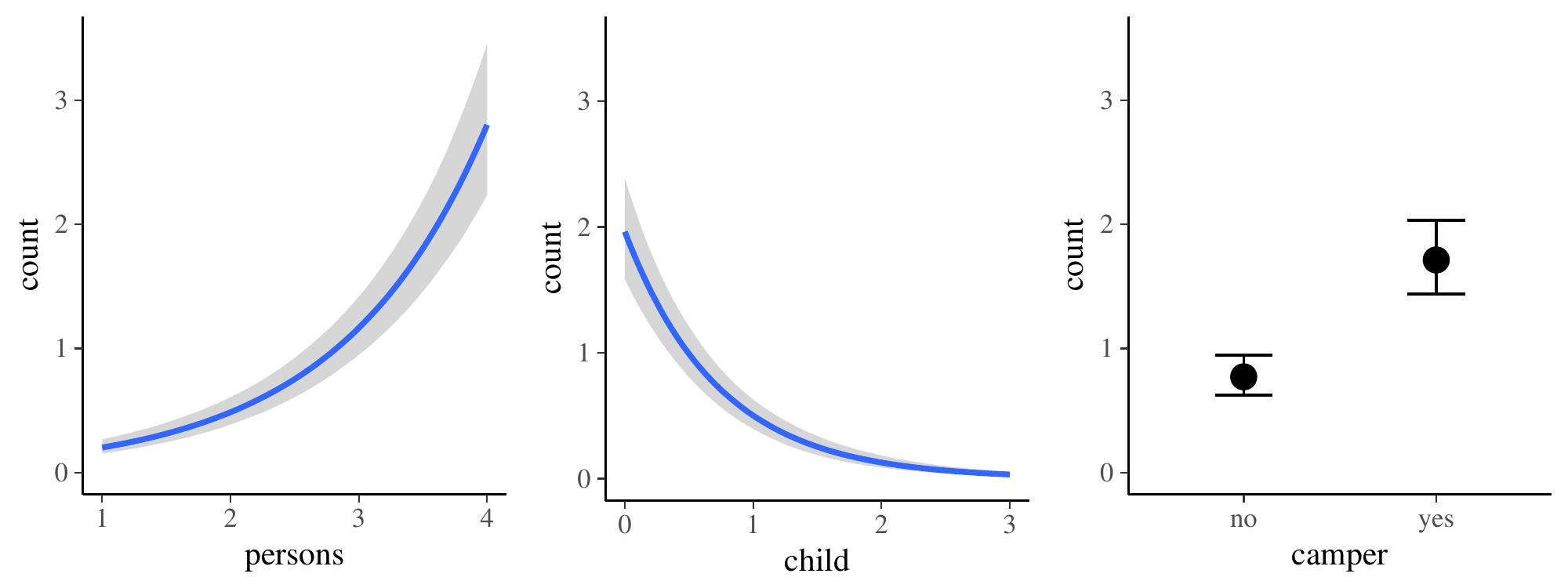}
	\caption{Marginal effects plots of the \code{fit\_zinb1} model.}
	\label{me_zinb1}
\end{figure}
(see Figure \ref{me_zinb1}). In fact, the \code{marginal\_effects} method turned out to be so powerful in visualizing effects of predictors that I am using it almost as frequently as \code{summary}. According to the parameter estimates, larger groups catch more fish, campers catch more fish than non-campers, and groups with more children catch less fish. The zero-inflation probability \code{zi} is pretty large with a mean of 41\%. Please note that the probability of catching no fish is actually higher than 41\%, but parts of this probability are already modeled by the Poisson distribution itself (hence the name zero-\emph{inflation}). If you want to treat all zeros as originating from a separate process, you can use hurdle models instead (not shown here). 

Now, we try to additionally predict the zero-inflation probability by the number of children. The underlying reasoning is that we expect groups with more children to not even try catching fish, since children often lack the patience required for fishing. From a purely statistical perspective, zero-inflated (and hurdle) distributions are a mixture of two processes and predicting both parts of the model is natural and often very reasonable to make full use of the data.

\begin{example}
fit_zinb2 <- brm(bf(count ~ persons + child + camper, zi ~ child), 
                 data = zinb, family = zero_inflated_poisson())
\end{example}
To transform the linear predictor of \code{zi} into a probability, \pkg{brms} applies the logit-link, which takes values within $[0, 1]$ and returns values on the real line. Thus, it allows the transition between probabilities and linear predictors. 

\begin{example}
summary(fit_zinb2)
\end{example}

\begin{example}
 Family: zero_inflated_poisson (log) 
Formula: count ~ persons + child + camper 
         zi ~ child
   Data: zinb (Number of observations: 250) 
Samples: 4 chains, each with iter = 2000; warmup = 1000; thin = 1; 
         total post-warmup samples = 4000
   WAIC: Not computed
 
Population-Level Effects: 
             Estimate Est.Error l-95
Intercept       -1.07      0.18    -1.43    -0.73       2322    1
persons          0.89      0.05     0.80     0.98       2481    1
child           -1.17      0.10    -1.37    -1.00       2615    1
camper           0.78      0.10     0.60     0.96       3270    1
zi_Intercept    -0.95      0.27    -1.52    -0.48       2341    1
zi_child         1.21      0.28     0.69     1.79       2492    1

Samples were drawn using sampling(NUTS). For each parameter, Eff.Sample 
is a crude measure of effective sample size, and Rhat is the potential 
scale reduction factor on split chains (at convergence, Rhat = 1).
\end{example}

According to the model, trying to fish with children not only decreases the overall number fish caught (as implied by the Poisson part of the model) but also drastically increases your chance of catching no fish at all (as implied by the zero-inflation part), possibly because groups with more children spend less time or no time at all fishing. Comparing model fit via leave-one-out cross validation as implemented in the \CRANpkg{loo} package \citep{loo2016, vehtari2016}.

\begin{example}
LOO(fit_zinb1, fit_zinb2)
\end{example}

\begin{example}
                        LOOIC     SE
fit_zinb1             1639.52 363.30
fit_zinb2             1621.35 362.39
fit_zinb1 - fit_zinb2   18.16  15.71
\end{example}
reveals that the second model using the number of children to predict both model parts has better fit. However, when considering  the standard error of the \code{LOOIC} difference, improvement in model fit is apparently modest and not substantial. More examples of distributional model can be found in \code{vignette("brms\_distreg")}.

\subsection{Example 2: Housing rents}

In their book about regression modeling, \citet{fahrmeir2013} use an example about the housing rents in Munich from 1999. The data contains information about roughly 3000 apartments including among others the absolute rent (\code{rent}), rent per square meter (\code{rentsqm}), size of the apartment (\code{area}), construction year (\code{yearc}), and the district in Munich (\code{district}), where the apartment is located. The data can be found in the \CRANpkg{gamlss.data} package \citep{gamlss.data}:

\begin{example}
data("rent99", package = "gamlss.data")
head(rent99)
\end{example}

\begin{example}
      rent   rentsqm area yearc location bath kitchen cheating district
1 109.9487  4.228797   26  1918        2    0       0        0      916
2 243.2820  8.688646   28  1918        2    0       0        1      813
3 261.6410  8.721369   30  1918        1    0       0        1      611
4 106.4103  3.547009   30  1918        2    0       0        0     2025
5 133.3846  4.446154   30  1918        2    0       0        1      561
6 339.0256 11.300851   30  1918        2    0       0        1      541
\end{example}
Here, we aim at predicting the rent per square meter with the size of the apartment as well as the construction year, while taking the district of Munich into account. As the effect of both predictors on the rent is of unknown non-linear form, we model these variables using a bivariate tensor spline \citep{wood2013}. The district is accounted for via a varying intercept.

\begin{example}
fit_rent1 <- brm(rentsqm ~ t2(area, yearc) + (1|district), data = rent99,
                 chains = 2, cores = 2)
\end{example}
We fit the model using just two chains (instead of the default of four chains) on two processor cores to reduce the model fitting time for the purpose of the present paper. In general, using the default option of four chains (or more) is recommended.

\begin{example}
summary(fit_rent1)
\end{example}

\begin{figure}[ht]
  \centering
  \includegraphics[width=0.99\textwidth,keepaspectratio]{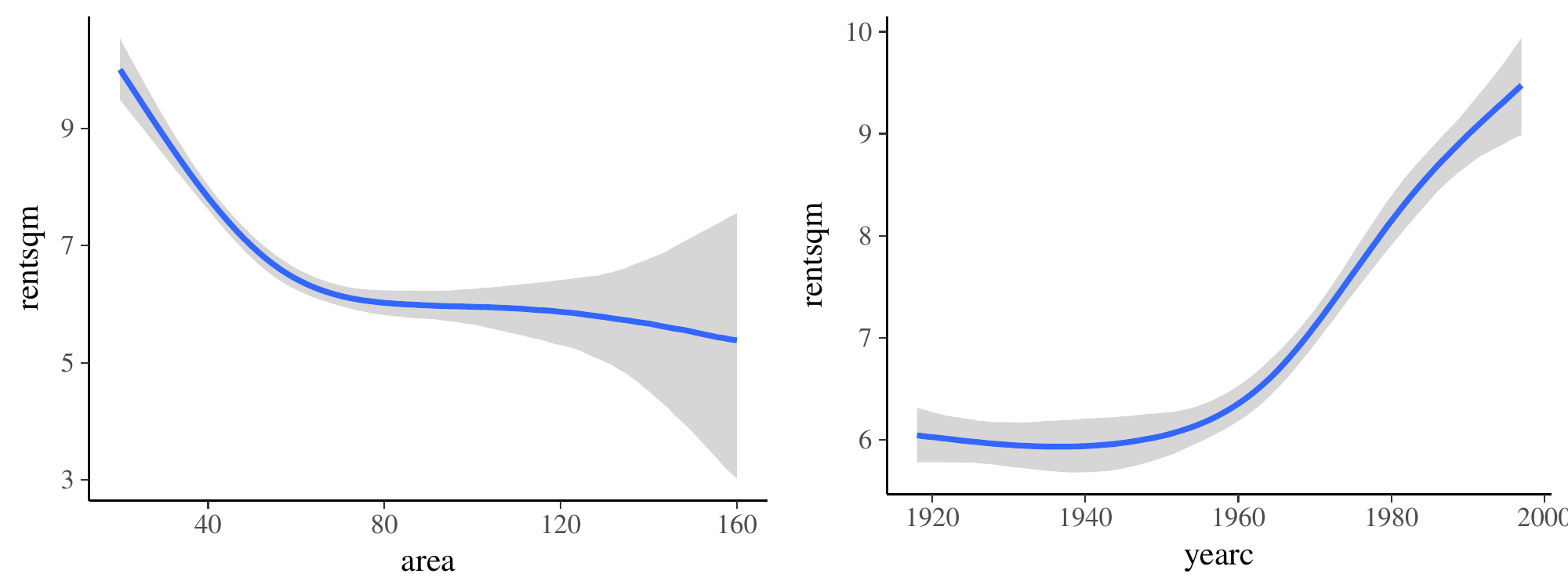}
	\caption{Marginal effects plots of the \code{fit\_rent1} model for single predictors.}
	\label{me_rent1}
\end{figure}

\begin{figure}[ht]
  \centering
  \includegraphics[width=0.7\textwidth,keepaspectratio]{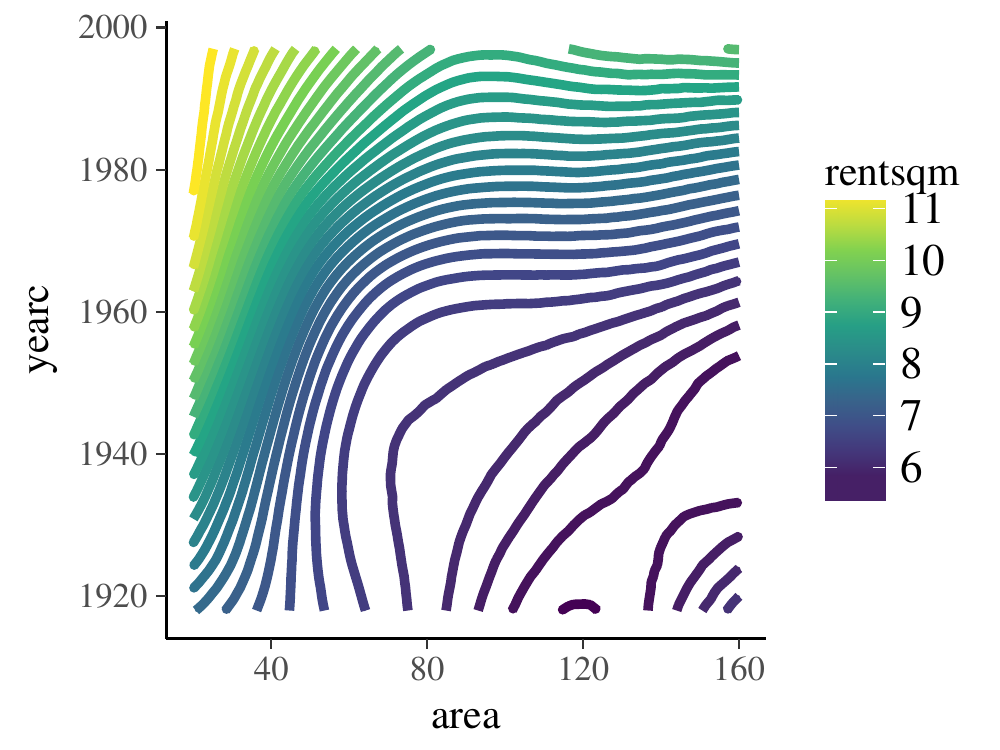}
	\caption{Surface plot of the \code{fit\_rent1} model for the combined effect of \code{area} and \code{yearc}.}
	\label{me_rent2}
\end{figure}

\begin{example}
 Family: gaussian(identity) 
Formula: rentsqm ~ t2(area, yearc) + (1 | district) 
   Data: rent99 (Number of observations: 3082) 
Samples: 2 chains, each with iter = 2000; warmup = 1000; thin = 1; 
         total post-warmup samples = 2000
    ICs: LOO = NA; WAIC = NA; R2 = NA
 
Smooth Terms: 
                   Estimate Est.Error l-95
sds(t2areayearc_1)     4.93      2.32     1.61    10.77       1546 1.00
sds(t2areayearc_2)     5.78      2.87     1.58    13.15       1175 1.00
sds(t2areayearc_3)     8.09      3.19     3.66    16.22       1418 1.00

Group-Level Effects: 
~district (Number of levels: 336) 
              Estimate Est.Error l-95
sd(Intercept)     0.60      0.06     0.48     0.73        494 1.01

Population-Level Effects: 
              Estimate Est.Error l-95
Intercept         7.80      0.11     7.59     8.02       2000 1.00
t2areayearc_1    -1.00      0.09    -1.15    -0.83       2000 1.00
t2areayearc_2     0.75      0.17     0.43     1.09       2000 1.00
t2areayearc_3    -0.07      0.16    -0.40     0.24       1579 1.00

Family Specific Parameters: 
      Estimate Est.Error l-95
sigma     1.95      0.03     1.90     2.01       2000 1.00

Samples were drawn using sampling(NUTS). For each parameter, Eff.Sample 
is a crude measure of effective sample size, and Rhat is the potential 
scale reduction factor on split chains (at convergence, Rhat = 1).
\end{example}
For models including splines, the output of \code{summary} is not tremendously helpful, but we get at least some information. Firstly, the credible intervals of the standard deviations of the coefficients forming the splines (under \code{'Smooth Terms'}) are sufficiently far away from zero to indicate non-linearity in the (combined) effect of \code{area} and \code{yearc}. Secondly, even after controlling for these predictors, districts still vary with respect to rent per square meter by a sizable amount as visible under \code{'Group-Level Effects'} in the output. To further understand the effect of the predictor, we apply graphical methods:

\begin{example}
marginal_effects(fit_rent1, surface = TRUE)
\end{example}
In Figure \ref{me_rent1}, the marginal effects of both predictors are displayed, while the respective other predictor is fixed at its mean. In Figure \ref{me_rent2}, the combined effect is shown, clearly demonstrating an interaction between the variables. In particular, housing rents appear to be highest for small and relatively new apartments.

In the above example, we only considered the mean of the response distribution to vary by \code{area} and \code{yearc}, but this my not necessarily reasonable assumption, as the variation of the response might vary with these variables as well. Accordingly, we fit splines and effects of district for both the location and the scale parameter, which is called \code{sigma} in Gaussian models.

\begin{figure}[ht]
  \centering
  \includegraphics[width=0.99\textwidth,keepaspectratio]{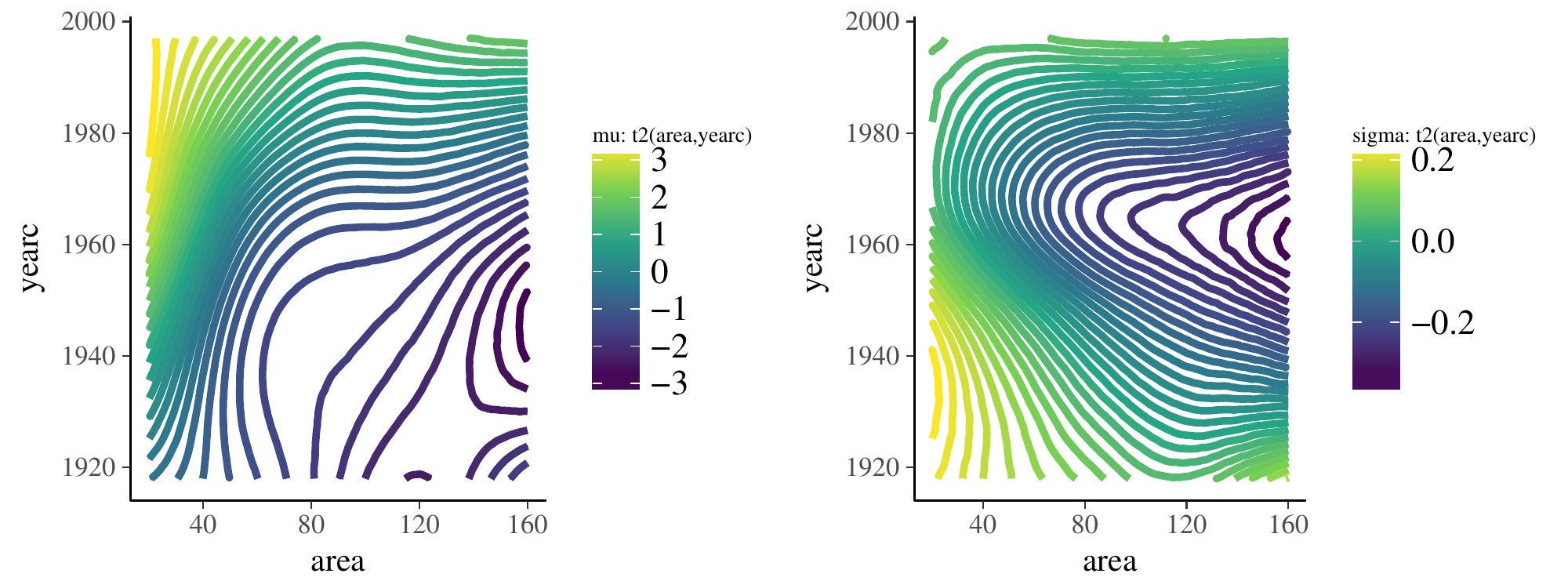}
	\caption{Plots showing the smooth terms of the \code{fit\_rent2} model.}
	\label{me_rent3}
\end{figure}

\begin{example}
bform <- bf(rentsqm ~ t2(area, yearc) + (1|ID1|district),
            sigma ~ t2(area, yearc) + (1|ID1|district))
fit_rent2 <- brm(bform, data = rent99, chains = 2, cores = 2)
\end{example}
If not otherwise specified, \code{sigma} is predicted on the log-scale to ensure it is positive no matter how the predictor term looks like. Instead of \code{(1|district)} as in the previous model, we now use \code{(1|ID1|district)} in both formulas. This results in modeling the varying intercepts of both model parts as correlated (see the description of the ID-syntax above). The group-level part of the \code{summary} output looks as follows:

\begin{example}
Group-Level Effects: 
~district (Number of levels: 336) 
                               Estimate Est.Error l-95
sd(Intercept)                      0.60      0.06     0.49     0.73        744 1.00
sd(sigma_Intercept)                0.11      0.02     0.06     0.15        751 1.00
cor(Intercept,sigma_Intercept)     0.72      0.17     0.35     0.98        648 1.00
\end{example}
As visible from the positive correlation of the intercepts, districts with overall higher rent per square meter have higher variation at the same time. Lastly, we want to turn our attention to the splines. While \code{marginal\_effects} is used to visualize effects of predictors on the expected response, \code{marginal\_smooths} is used to show just the spline parts of the model:

\begin{example}
marginal_smooths(fit_rent2)
\end{example} 
The plot on the left-hand side of Figure \ref{me_rent3} resembles the one in Figure \ref{me_rent2}, but the scale is different since only the spline is plotted. The right-hand side of \ref{me_rent3} shows the spline for \code{sigma}. Since we apply the log-link on \code{sigma} by default the spline is on the log-scale as well. As visible in the plot, the variation in the rent per square meter is highest for relatively small and old apartments, while the variation is smallest for medium to large apartments build around the 1960s.

\subsection{Example 3: Insurance loss payments}

On his blog, Markus Gesmann predicts the growth of cumulative insurance loss payments over time, originated from different origin years (see \url{http://www.magesblog.com/2015/11/loss-developments-via-growth-curves-and.html}). We will use a slightly simplified version of his model for demonstration purposes here. It looks as follows:

$$cum_{AY, dev} \sim N(\mu_{AY, dev}, \sigma)$$
$$\mu_{AY, dev} = ult_{AY} \left(1 - \exp\left(- \left( \frac{dev}{\theta} \right)^\omega \right) \right)$$

The cumulative insurance payments $cum$ will grow over time, and we model this dependency using the variable $dev$. Further, $ult_{AY}$ is the (to be estimated) ultimate loss of accident each year. It constitutes a non-linear parameter in our framework along with the parameters $\theta$ and $\omega$, which are responsible for the growth of the cumulative loss and are for now assumed to be the same across years. We load the data

\begin{example}
url <- paste0("https://raw.githubusercontent.com/mages/",
              "diesunddas/master/Data/ClarkTriangle.csv")
loss <- read.csv(url)
head(loss)
\end{example}

\begin{example}
    AY dev      cum
1 1991   6  357.848
2 1991  18 1124.788
3 1991  30 1735.330
4 1991  42 2182.708
5 1991  54 2745.596
6 1991  66 3319.994
\end{example}
and translate the proposed model into a non-linear \pkg{brms} model.

\begin{example}
nlform <- bf(cum ~ ult * (1 - exp(-(dev / theta)^omega)),
             ult ~ 1 + (1|AY), omega ~ 1, theta ~ 1, nl = TRUE)
             
nlprior <- c(prior(normal(5000, 1000), nlpar = "ult"),
             prior(normal(1, 2), nlpar = "omega"),
             prior(normal(45, 10), nlpar = "theta"))
              
fit_loss1 <- brm(formula = nlform, data = loss, family = gaussian(), 
                 prior = nlprior, control = list(adapt_delta = 0.9))
\end{example}

In the above functions calls, quite a few things are going on. The formulas are wrapped in \code{bf} to combine them into one object. The first formula specifies the non-linear model. We set argument \code{nl = TRUE} so that \pkg{brms} takes this formula literally and instead of using standard R formula parsing. We specify one additional formula per non-linear parameter (a) to clarify what variables are covariates and what are parameters and (b) to specify the predictor term for the parameters. We estimate a group-level effect of accident year (variable \code{AY}) for the ultimate loss \code{ult}. This also shows nicely how a non-linear parameter is actually a placeholder for a linear predictor, which in the case of \code{ult}, contains only a varying intercept for year. Both \code{omega} and \code{theta} are assumed to be constant across observations so we just fit a population-level intercept. 

Priors on population-level effects are required and, for the present model, are actually mandatory to ensure identifiability. Otherwise, we may observe that different Markov chains converge to different parameter regions as multiple posterior distribution are equally plausible. Setting prior distributions is a difficult task especially in non-linear models. It requires some experience and knowledge both about the model that is being fitted and about the data at hand. Additionally, there is more to keep in mind to optimize the sampler's performance: Firstly, using non- or weakly informative priors in non-linear models often leads to problems even if the model is generally identified. For instance, if a zero-centered and reasonably wide prior such as \code{normal(0, 10000)} it set on \code{ult}, there is little information about \code{theta} and \code{omega} for samples of \code{ult} being close to zero, which may lead to convergence problems. Secondly, Stan works best when parameters are roughly on the same order of magnitude \citep{stan2017}. In the present example, \code{ult} is of three orders larger than \code{omega}. Still, the sampler seems to work quite well, but this may not be true for other models. One solution is to rescale parameters before model fitting. For instance, for the present example, one could have downscaled \code{ult} by replacing it with \code{ult * 1000} and correspondingly the \code{normal(5000, 1000)} prior with \code{normal(5, 1)}.

In the \code{control} argument we increase \code{adapt\_delta} to get rid of a few divergent transitions (cf. \citeauthor{stan2017}, \citeyear{stan2017}; \citeauthor{buerkner2017}, \citeyear{buerkner2017}). Again the model is summarized via

\begin{example}
summary(fit_loss1)
\end{example}

\begin{example}
 Family: gaussian (identity) 
Formula: cum ~ ult * (1 - exp(-(dev / theta)^omega)) 
         ult ~ 1 + (1 | AY)
         omega ~ 1
         theta ~ 1
   Data: loss (Number of observations: 55) 
Samples: 4 chains, each with iter = 2000; warmup = 1000; thin = 1; 
         total post-warmup samples = 4000
   WAIC: Not computed
 
Group-Level Effects: 
~AY (Number of levels: 10) 
                  Estimate Est.Error l-95
sd(ult_Intercept)   745.74    231.31   421.05  1306.04        916    1

Population-Level Effects: 
                Estimate Est.Error l-95
ult_Intercept    5273.70    292.34  4707.11  5852.28        798    1
omega_Intercept     1.34      0.05     1.24     1.43       2167    1
theta_Intercept    46.07      2.09    42.38    50.57       1896    1

Family Specific Parameters: 
      Estimate Est.Error l-95
sigma   139.93     15.52    113.6   175.33       2358    1

Samples were drawn using sampling(NUTS). For each parameter, Eff.Sample 
is a crude measure of effective sample size, and Rhat is the potential 
scale reduction factor on split chains (at convergence, Rhat = 1).
\end{example}
as well as

\begin{example}
marginal_effects(fit_loss1)
\end{example}
\begin{figure}[ht]
  \centering
  \includegraphics[width=0.7\textwidth,keepaspectratio]{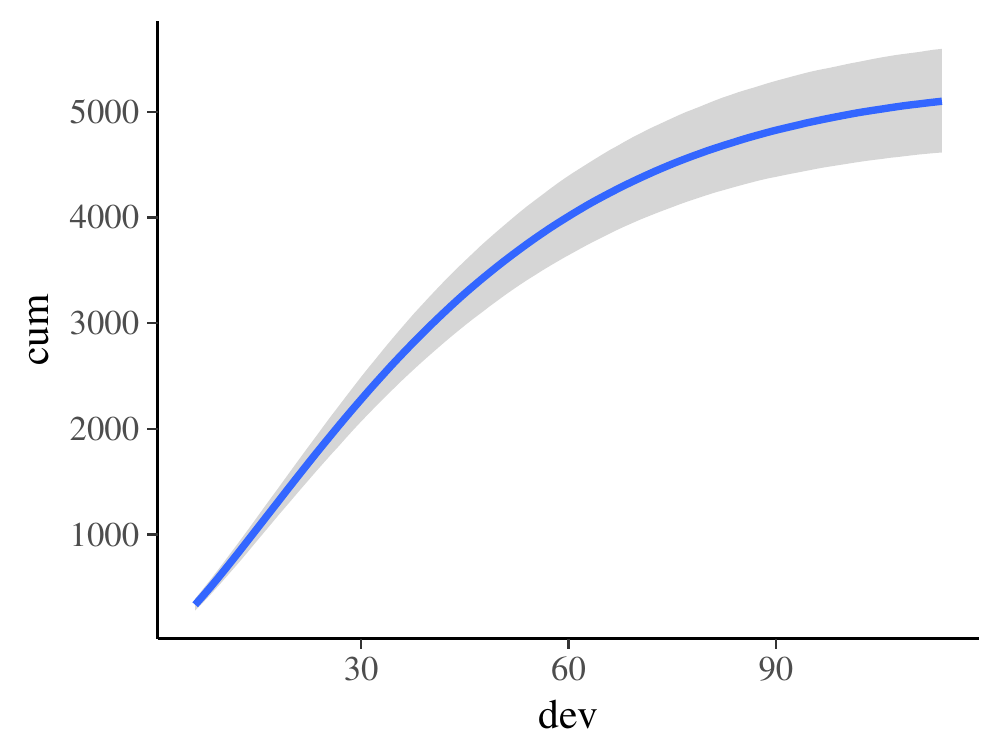}
	\caption{Marginal effects plots of the \code{fit\_loss1} model.}
	\label{me_loss1}
\end{figure}
(see Figure \ref{me_loss1}). We can also visualize the cumulative insurance loss over time separately for each year.

\begin{example}
conditions <- data.frame(AY = unique(loss$AY))
rownames(conditions) <- unique(loss$AY)
me_year <- marginal_effects(fit_loss1, conditions = conditions, 
                            re_formula = NULL, method = "predict")
plot(me_year, ncol = 5, points = TRUE)
\end{example}
\begin{figure}[ht]
  \centering
  \includegraphics[width=0.99\textwidth,keepaspectratio]{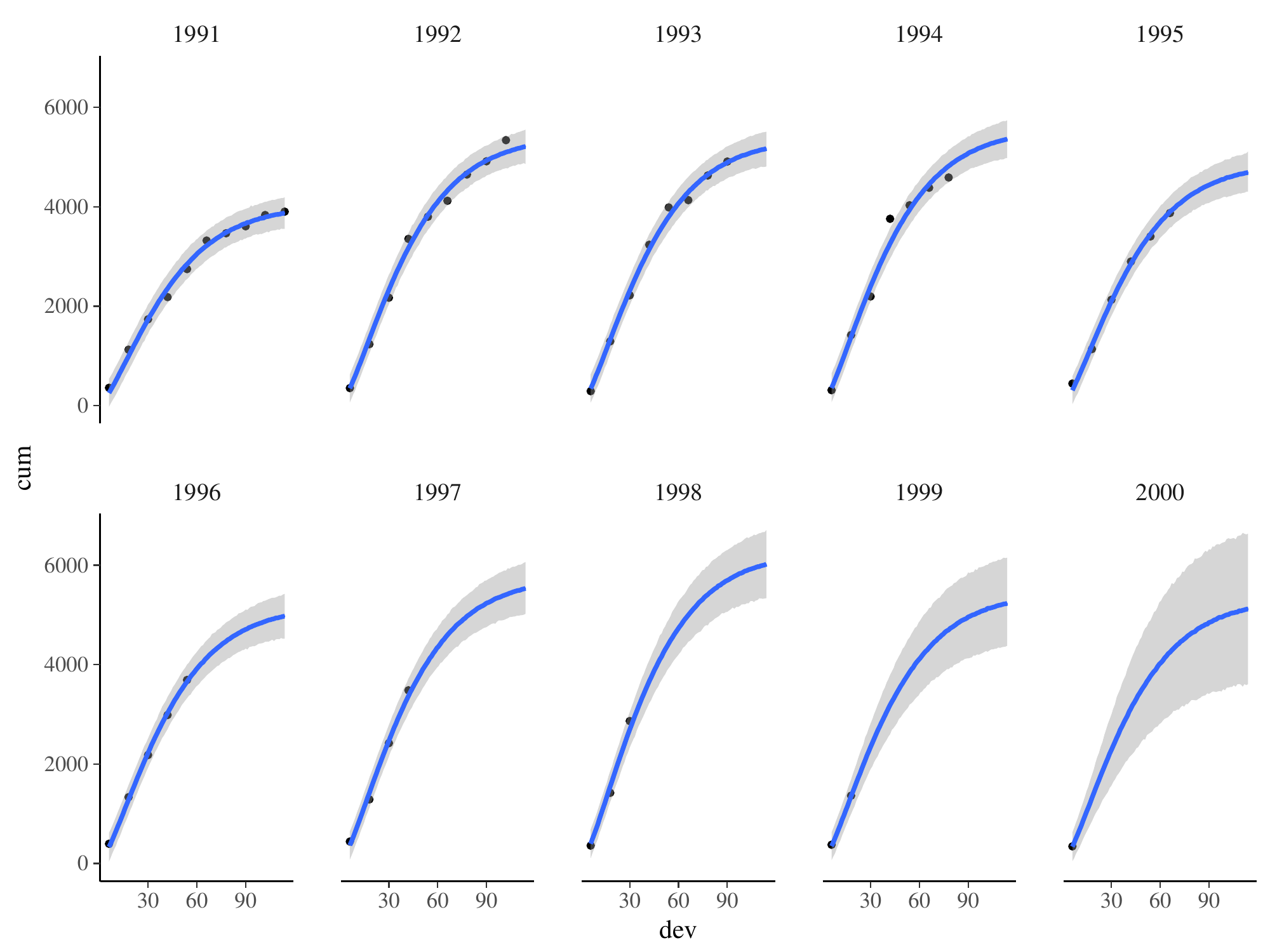}
	\caption{Marginal effects plots of the \code{fit\_loss1} model separately for each accident year.}
	\label{me_loss1_year}
\end{figure}
(see Figure \ref{me_loss1_year}). It is evident that there is some variation in cumulative loss across accident years, for instance due to natural disasters happening only in certain years. Further, we see that the uncertainty in the predicted cumulative loss is larger for later years with fewer available data points. 

In the above model, we considered \code{omega} and \code{delta} to be constant across years, which may not necessarily be true. We can easily investigate this by fitting varying intercepts for all three non-linear parameters also estimating group-level correlation using the above introduced \code{ID} syntax.

\begin{example}
nlform2 <- bf(cum ~ ult * (1 - exp(-(dev / theta)^omega)),
              ult ~ 1 + (1|ID1|AY), omega ~ 1 + (1|ID1|AY),
              theta ~ 1 + (1|ID1|AY), nl = TRUE)
              
fit_loss2 <- update(fit_loss1, formula = nlform2,
                    control = list(adapt_delta = 0.90))
\end{example}
We could have also specified all predictor terms more conveniently within one formula as 
\begin{example}
ult + omega + theta ~ 1 + (1|ID1|AY)
\end{example}
because the structure of the predictor terms is identical. To compare model fit, we perform leave-one-out cross-validation.

\begin{example}
LOO(fit_loss1, fit_loss2)
\end{example}

\begin{example}
                      LOOIC    SE
fit_loss1             715.44 19.24
fit_loss2            720.60 19.85
fit_loss1 - fit_loss2  -5.15  5.34
\end{example}

Since smaller values indicate better expected out-of-sample predictions and thus better model fit, the simpler model that only has a varying intercept over parameter \code{ult} is preferred. This may not be overly surprising, given that three varying intercepts as well as three group-level correlations are probably overkill for data containing only 55 observations. Nevertheless, it nicely demonstrates how to apply the \code{ID} syntax in practice. More examples of non-linear models can be found in \code{vignette("brms\_nonlinear")}.

\subsection{Example 4: Performance of school children}

Suppose that we want to predict the performance of students in the final exams at the end of the year. There are many variables to consider, but one important factor will clearly be school membership. Schools might differ in the ratio of teachers and students, the general quality of teaching, in the cognitive ability of the students they draw, or other factors we are not aware of that induce dependency among students of the same school. Thus, it is advised to apply a multilevel modeling techniques including school membership as a group-level term. Of course, we should account for class membership and other levels of the educational hierarchy as well, but for the purpose of the present example, we will focus on schools only. Usually, accounting for school membership is pretty-straight forward by simply adding a varying intercept to the formula: \code{(1 | school)}. However, a non-negligible number of students might change schools during the year. This would result in a situation where one student is a member of multiple schools and so we need a multi-membership model. Setting up such a model not only requires information on the different schools students attend during the year, but also the amount of time spend at each school. The latter can be used to weight the influence each school has on its students, since more time attending a school will likely result in greater influence. For now, let us assume that students change schools maximally once a year and spend equal time at each school. We will later see how to relax these assumptions.

Real educational data are usually relatively large and complex so that we simulate our own data for the purpose of this tutorial paper. We simulate 10 schools and 1000 students, with each school having the same expected number of 100 students. We model 10\% of students as changing schools.

\begin{example}
data_mm <- sim_multi_mem(nschools = 10, nstudents = 1000, change = 0.1)
head(data_mm)
\end{example}

\begin{example}
  s1 s2  w1  w2        y
1  8  9 0.5 0.5 16.27422
2 10  9 0.5 0.5 18.71387
3  5  3 0.5 0.5 23.65319
4  3  5 0.5 0.5 22.35204
5  5  3 0.5 0.5 16.38019
6 10  6 0.5 0.5 17.63494
\end{example}
The code of function \code{sim\_multi\_mem} can be found in the online supplement of the present paper. For reasons of better illustration, students changing schools appear in the first rows of the data. Data of students being only at a single school looks as follows:

\begin{example}
data_mm[101:106, ]
\end{example}

\begin{example}
    s1 s2  w1  w2         y
101  2  2 0.5 0.5 27.247851
102  9  9 0.5 0.5 24.041427
103  4  4 0.5 0.5 12.575001
104  2  2 0.5 0.5 21.203644
105  4  4 0.5 0.5 12.856166
106  4  4 0.5 0.5  9.740174
\end{example}
Thus, school variables are identical, but we still have to specify both in order to pass the data appropriately. Incorporating no other predictors into the model for simplicity, a multi-membership model is specified as

\begin{example}
fit_mm <- brm(y ~ 1 + (1 | mm(s1, s2)), data = data_mm)
\end{example}
The only new syntax element is that multiple grouping factors (\code{s1} and \code{s2}) are wrapped in \code{mm}. Everything else remains exactly the same. Note that we did not specify the relative weights of schools for each student and thus, by default, equal weights are assumed. 

\begin{example}
summary(fit_mm)
\end{example}

\begin{example}
 Family: gaussian (identity) 
Formula: y ~ 1 + (1 | mm(s1, s2)) 
   Data: data_mm (Number of observations: 1000) 
Samples: 4 chains, each with iter = 2000; warmup = 1000; thin = 1; 
         total post-warmup samples = 4000
   WAIC: Not computed
 
Group-Level Effects: 
~mms1s2 (Number of levels: 10) 
              Estimate Est.Error l-95
sd(Intercept)     2.76      0.82     1.69     4.74        682 1.01

Population-Level Effects: 
          Estimate Est.Error l-95
Intercept       19      0.93    17.06     20.8        610    1

Family Specific Parameters: 
      Estimate Est.Error l-95
sigma     3.58      0.08     3.43     3.75       2117    1

Samples were drawn using sampling(NUTS). For each parameter, Eff.Sample 
is a crude measure of effective sample size, and Rhat is the potential 
scale reduction factor on split chains (at convergence, Rhat = 1).
\end{example}

With regard to the assumptions made in the above example, it is unlikely that all children who change schools stay in both schools equally long. To relax this assumption, we have to specify weights. First, we amend the simulated data to contain non-equal weights for students changing schools. For all other students, weighting does of course not matter as they stay in the same school anyway.

\begin{example}
data_mm[1:100, "w1"] <- runif(100, 0, 1)
data_mm[1:100, "w2"] <- 1 - data_mm[1:100, "w1"]
head(data_mm)
\end{example}

\begin{example}
  s1 s2        w1         w2        y
1  8  9 0.3403258 0.65967423 16.27422
2 10  9 0.1771435 0.82285652 18.71387
3  5  3 0.9059811 0.09401892 23.65319
4  3  5 0.4432007 0.55679930 22.35204
5  5  3 0.8052026 0.19479738 16.38019
6 10  6 0.5610243 0.43897567 17.63494
\end{example}
Incorporating these weights into the model is straight forward.

\begin{example}
fit_mm2 <- brm(y ~ 1 + (1 | mm(s1, s2, weights = cbind(w1, w2))), 
               data = data_mm)
\end{example}
The summary output is similar to the previous, so we do not show it here. The second assumption that students change schools only once a year, may also easily be relaxed by providing more than two grouping factors, say, \code{mm(s1, s2, s3)}.

\section{Conclusion}

The present paper is meant to introduce R users and developers to the extended \pkg{lme4} formula syntax applied in \pkg{brms}. Only a subset of modeling options were discussed in detail, which ensured the paper was not too broad. For some of the more basic models that \pkg{brms}  can fit, see \citet{buerkner2017}. Many more examples can be found in the growing number of vignettes accompanying the package (see \code{vignette(package = "brms")} for an overview). 

To date, \pkg{brms} is already one of the most flexible R packages when it comes to regression modeling. However, for the future, there are quite a few more features that I am planning to implement (see \url{https://github.com/paul-buerkner/brms/issues} for the current list of issues). In addition to smaller, incremental updates, I have two specific features in mind: extended multivariate models and missing value imputation. I receive ideas and suggestions from users almost every day -- for which I am always grateful -- and so the list of features that will be implemented in the proceeding versions of \pkg{brms} will continue to grow. 

\section*{Acknowledgments}

First of all, I would like to thank the Stan Development Team for creating the probabilistic programming language Stan, which is an incredibly powerful and flexible tool for performing full Bayesian inference. Without it, \pkg{brms} could not fit a single model. Furthermore, I want to thank Heinz Holling, Donald Williams and Ruben Arslan for valuable comments on earlier versions of the paper. I also would like to thank the many users who reported bugs or had ideas for new features, thus helping to continuously improve \pkg{brms}. 

\bibliography{brms_multilevel}

\address{Paul-Christian B\"urkner \\
  Faculty of Psychology, University of M\"unster \\
  Fliednerstr. 21, 48151 M\"unster \\
  Germany\\
  \email{paul.buerkner@gmail.com}
 }

\end{article}

\end{document}